\documentclass[twocolumn,nofootinbib,prd,aps,superscriptaddress,tightenlines]{revtex4-1}

\usepackage{slashed}
\usepackage{braket}
\usepackage{bm}

\newcommand{\nn}{\nonumber \\ }
\def\vev#1{\left\langle #1 \right\rangle}
\def\abs#1{\left| #1 \right|}
\def\rd{{\rm d}}
\def\lqcd{\Lambda_{\text{QCD}}}

\begin{document}

\title{Reparametrization Invariance Constraints on Inclusive Decay Spectra and Masses}

\author{Aneesh V.~Manohar}
\affiliation{Department of Physics, University of California at San Diego,
  La Jolla, CA 92093}

\begin{abstract}
Reparametrization invariance relations are determined for the inclusive decay spectra of heavy hadrons. They connect the leading order result to the $\lambda_1$ term, and the $\lambda_2$ and $\rho_2$ terms, and imply that the $\lambda_1$ and $\rho_2$ terms in the total rate occur in the combinations $1+\lambda_1/(2m_Q^2)$ and $\lambda_2-\rho_2/m_Q$.
The relations are satisfied by the known results for the hadronic decay tensors for $\bar B \to X_{c,u} \ell \nu$, $\bar B \to X_s \gamma$ and $\bar B \to X_s \ell^+ \ell^-$ decays.
An interesting field-theory result, the connection between currents in full and effective theories and the change in currents under field redefinitions, is discussed. The hadron masses are given to order $\lqcd^3/m_Q^2$ including radiative corrections, and provide an example illustrating the relation between renormalization and reparametrization invariance and $\mu$-independence of $1/m_Q^3$ corrections to physical quantities.
\end{abstract}

\date{\today}

\maketitle

\section{Introduction}\label{sec:intro}

Heavy quark effective theory (HQET) is an effective field theory that describes the interactions of heavy quarks in a systematic expansion in inverse powers of the quark mass $m_Q$. The effective field theory (EFT) depends on a four-velocity $v^\mu$, and has quark fields with velocity labels $Q_v$~\cite{hqetbook}. The total momentum of a heavy quark is $p=m_Q v + k$, where $k$ is referred to as the residual momentum. HQET describes heavy quark interactions as long as the residual momentum $k$ is parametrically smaller than $m_Q$; in most applications the residual momentum $k$ is of order $\lqcd$, and the HQET expansion parameter is $\lqcd/m_Q$.

The split of the total momentum $p$ into $m_Q v$ and $k$ is not unique. One can redefine the velocity of the heavy quark by the transformation $k \to k - l$, $v \to v + l/m_Q$, where $l$ is of order $\lqcd$. This transformation, which leaves the total momentum unchanged and preserves the HQET power counting, is called reparametrization invariance~\cite{Luke:1992cs}. Reparametrization invariance (RPI) leads to important relations in the effective theory~\cite{Luke:1992cs,Manohar:1997qy} between terms at different orders in $1/m_Q$. In this paper, we derive the consequences of reparametrization invariance for inclusive decay spectra to order $1/m_Q^3$, and verify that they hold for existing computations of the decay spectra for $\bar B \to X_{c,u} \ell \nu$, $\bar B \to X_s \gamma$ and $\bar B \to X_s \ell^+ \ell^-$ decay to order $1/m_Q^3$~\cite{Chay:1990da,[][{ [Erratum, {\sl ibid.}\ {\bf B297}, 477 (1993)]}]Bigi:1992su,Bigi:1993fe,Manohar:1993qn,Gremm:1996df,Falk:1993dh,Ali:1996bm,[][{ [Erratum, {\sl ibid.}\ {\bf D60}, 099907 (1999)]}]Bauer:1997fe,Bauer:1999kf}. RPI relates the $\lambda_1$ terms in the decay spectra to the leading order result; this connection has been noted previously for the total rate~\cite{Bigi:1992su} and for the triple differential rate~\cite{Manohar:1993qn}. RPI also relates the $\lambda_2$ and $\rho_2$ terms in the decay spectra and in the total rate.

An interesting field-theoretic point --- the relation between full and effective theory currents, and the change in currents under field redefinitions  is discussed in detail in Sec.~\ref{sec:currents}. Hadron masses to order $\lqcd^3/m_Q^2$ including radiative corrections are given in the Appendix. The mass formul\ae\ are RPI invariant, and the renormalization group (RG) evolution respects RPI. The masses have a contribution from four-quark operators, which is necessary for them to be $\mu$-independent.

\section{The Lagrangian}\label{sec:lag}

The HQET Lagrangian is constructed to reproduce the on-shell scattering amplitudes ($S$-matrix elements) of the original QCD theory.
The HQET Lagrangian is not unique --- different Lagrangians related by a field redefinition are equivalent, and lead to the same physical predictions for all measurable quantities. It is convenient to choose a standard form for the HQET Lagrangian; we will pick the form in which time-derivatives of the heavy quark field other than the leading $i D \cdot v$ term are eliminated by making field redefinitions.

The most general effective Lagrangian to order $1/m^2$ (up to field redefinitions) is
\begin{eqnarray}
{\cal L} &=& {\cal L}_{l} + \overline Q_v \Biggl\{ 
i D\cdot v - c_2\, \frac{D_\perp^2}{2 m_Q} - c_F\, g \frac{\sigma_{\alpha\beta} G^{\alpha\beta}}{4 m_Q} \nn
&&- c_D\, g \frac{v^{\alpha}\left[D^\beta_\perp  G_{\alpha\beta}\right]}{8 m_Q^2} + i c_S\, g \frac{v_\lambda
\sigma_{\alpha\beta}\left\{D^\alpha_\perp , G^{\lambda\beta}\right\}}{8 m_Q^2} \Biggr\} Q_v \nn
&& +\frac{1}{m_Q^2} \sum_i c_i^{(2)} O_i^{(2)}\,,
\label{eq1}
\end{eqnarray}
where the covariant derivative is $D^\mu = \partial^\mu + i g A^{\mu}$, and the $\perp$ component of a four-vector $V^\mu$ is defined by
\begin{eqnarray}
V_\perp^\mu &\equiv& g_\perp^{\mu \nu}V_\nu \,,\nn
g_\perp^{\mu \nu} &\equiv& g^{\mu \nu}-v^\mu v^\nu\,.
\end{eqnarray}
Covariant derivatives in square brackets act only
on the fields within the brackets. The other covariant derivatives act on all
fields to the right. The subscripts $F$, $S$ and $D$ stand for Fermi,
spin-orbit, and Darwin, respectively. The order $1/m^0$ terms in the Lagrangian involving only the light fields (light quarks and gluons) are denoted by ${\cal L}_l$. The operators $O_i^{(2)}$ are $1/m^2$ four-quark operators, penguin operators, and higher derivative gluon terms, and are listed in Ref.~\cite{Bauer:1997gs}.  It is convenient to define the operators
\begin{eqnarray}
O_2 &=& -\frac{1}{2m_Q}\,\overline Q_v D_\perp^2 \overline Q_v\,,\nn
O_F &=& -\frac{g}{4m_Q}\,\overline Q_v \sigma^{\alpha\beta}G_{\alpha\beta}Q_v\,,\nn
O_D &=& -\frac{gv^\alpha}{8m_Q^2}\,\overline Q_v \left[D_\perp^\beta, G_{\alpha\beta}\right] Q_v\,,\nn
O_S &=& \frac{i gv_\lambda }{8m_Q^2}\, \overline Q_v \sigma_{\alpha\beta}\left\{D_\perp^\alpha, G_{\lambda\beta}\right\}Q_v\,,
\label{ops}
\end{eqnarray}
so that the Lagrangian is
\begin{eqnarray}
{\cal L} &=& {\cal L}_{l} + \overline Q_v (i D \cdot v) Q_v +c_2 O_2 + c_F O_F\nn
&&+c_D O_D + c_S O_S +\frac{1}{m_Q^2} \sum_i c_i^{(2)} O_i^{(2)}\,.
\label{eq1alt}
\end{eqnarray}

The $B$ and $B^*$ meson states in the effective theory, $\ket{B^{(*)}\!\!,v}$ are defined as states constructed using the leading order ($ m \to \infty$) effective Lagrangian, and are normalized to $v^0$. With this convention, the matrix elements of the operators in Eq.~(\ref{ops}) are
\begin{eqnarray}
\braket{\bar B^{(*)}\!\!,v|O_2|\bar B^{(*)}\!\!,v} &=& \frac{\lambda_1}{2m_Q}\,,\nn
\braket{\bar B^{(*)}\!\!,v|O_F|\bar B^{(*)}\!\!,v}  &=& \frac{d_H \lambda_2}{2m_Q}\,,\nn
\braket{\bar B^{(*)}\!\!,v|O_D|\bar B^{(*)}\!\!,v} &=& - \frac{\rho_1}{4m_Q^2}\,,\nn
\braket{\bar B^{(*)}\!\!,v|O_S|\bar B^{(*)}\!\!,v}  &=& -\frac{d_H \rho_2}{4m_Q^2}\,,
\label{eq5}
\end{eqnarray}
where $d_H=3$ for $B$ mesons, and $d_H=-1$ for $B^*$ mesons. These equations define the non-perturbative parameters $\lambda_{1,2}$ of order $\lqcd^2$, and $\rho_{1,2}$ of order $\lqcd^3$.

Matrix elements of the operators between full theory states $\ket{\bar B^{(*)},p}$ have contributions of higher order in $1/m_Q$ from time-ordered products of the operators with subleading terms in the EFT Lagrangian, which are given by additional non-perturbative parameters $\mathcal{T}_{1-4}$~\cite{Gremm:1996df}.

\section{EFT Currents and Field Redefinitions}\label{sec:currents}

An important feature of the HQET computation of inclusive decay rates is that the leading term in the $1/m_Q$ expansion  is absolutely normalized, with no unknown non-perturbative matrix elements. This is because the leading term in the OPE can be written in terms of the full theory current $\bar b \gamma^\mu b$. The forward matrix element  of the full theory current $\bar b \gamma^\mu b$
between full theory states  is exactly unity, since the operator is the symmetry current of $b$-quark number, which is conserved in QCD.\footnote{There are some subtleties, see Ref.~\cite{Collins:2005nj}.} This is the procedure followed in the tree-level computations of Refs.~\cite{Chay:1990da,Bigi:1992su,Bigi:1993fe,Manohar:1993qn}.

The full theory field $b$ is expanded in terms of a field $b_v$,
\begin{eqnarray}
b &=& \left[1 + \frac{1}{2m+i v \cdot D} i \slashed{D}_\perp \right] b_v\,.
\label{eq6a}
\end{eqnarray}
Expanding out the vector and scalar operators in the full theory gives the relation
\begin{eqnarray}
\overline b b - \overline b \gamma^0 b &=& -\frac{1}{m_Q^2} \overline b_v D_\perp^2 b_v + \ldots
\label{eq7a}
\end{eqnarray}
which gives the well known tree-level result that the matrix element of $\overline b b$ is $1+\lambda_1/(2m_Q^2)$ up to order $1/m_Q^2$~\cite{Bigi:1992su}.\footnote{The OPE gives the decay rate in terms of 
the scalar operator $\overline b b$.} Unfortunately, Eq.~(\ref{eq6a}) is meaningless beyond tree-level since the full and effective theory are renormalized differently. We need a relation analogous to Eq.~(\ref{eq7a}) valid to all orders in $\alpha_s$, and to order $1/m_Q^3$, which allows us to compute EFT matrix elements with an absolute normalization.

In this paper, RPI is used to obtain results on the inclusive decay spectra to all orders in $\alpha_s$. To normalize the leading term would require computing the matching from the full theory current $\bar b \gamma^\mu b$ onto EFT operators to order $1/m_Q^3$. Instead, we can obtain the matrix element by using the fact that $b$-quark number is a symmetry of the EFT, so that the EFT current has unit matrix element. This avoids the necessity of a matching computation for the current to all orders in $\alpha_s$.
The current in the EFT determined from the Lagrangian Eq.~(\ref{eq1}) by the transformation
\begin{eqnarray}
Q_v &\to& e^{i \alpha } Q_v
\end{eqnarray}
is
\begin{eqnarray}
j^0 &=& \overline Q_v Q_v\,,
\label{current}
\end{eqnarray}
and has forward matrix element unity to all orders in $1/m$. In deriving Eq.~(\ref{current}), it is crucial that $\bar Q_v ( D \cdot v)Q_v$ is the only term with time derivatives of $Q_v$ in the Lagrangian. Suppose instead that the Lagrangian written in terms of some field $h_v$ had higher order time-derivative terms, for example
\begin{eqnarray}
{\cal L} ^\prime &=&{\cal L} + c \frac{g }{2 m_Q^2}\overline h_v \left\{iv \cdot D, \sigma_{\alpha \beta}G^{\alpha \beta}\right\}h_v\,,
\label{eq1a}
\end{eqnarray}
where ${\cal L}$ is the original Lagrangian Eq.~(\ref{eq1}) but with $Q_v \to h_v$, and $c$ is a constant. Then the current is
\begin{eqnarray}
\left(j^\prime\right)^0 &=& \overline h_v h_v + c \frac{g }{m_Q^2}\overline h_v \sigma_{\alpha \beta}G^{\alpha \beta}h_v\,,
\label{eq11a}
\end{eqnarray}
and it is $\left(j^\prime\right)^0$ whose forward matrix element is normalized to unity, so that
the matrix element of $\overline h_v h_v$ is $1-4 c \vev{O_F}/m_Q=1-2 d_H c \lambda_2/m_Q^2 + {\cal O}(1/m_Q^3)$.  Note that the additional term in the Lagrangian Eq.~(\ref{eq1a}) is a $i v \cdot D$ term, but the term in the current does not have a $i v \cdot D$ factor.
Generically, terms with $n$ time-derivatives in the Lagrangian will produce terms in the current with $n-1$ time derivatives.
The field redefinition
\begin{eqnarray}
h_v &=& \left[1-c \frac{g }{2 m_Q^2} \sigma_{\alpha \beta}G^{\alpha \beta}\right]Q_v
\end{eqnarray}
eliminates the $\left\{iv \cdot D, \sigma_{\alpha \beta}G^{\alpha \beta}\right\}$ term,
converts Eq.~(\ref{eq1a}) back into the standard form Eq.~(\ref{eq1}), and converts $\left(j^\prime\right)^0$ into $j^0$.

The EFT current $j^0$ is changed by field redefinitions. A simple example that demonstrates this is to consider a field redefinition involving time-derivatives,
\begin{eqnarray}
\widetilde Q_v &=& Q_v + \frac{d}{m_Q} (i D \cdot v) Q_v\,,
\label{eq53}
\end{eqnarray}
where $d$ is a constant. The theory is described by the Lagrangians ${\cal L}(Q_v)$ or by $\widetilde {\cal L}(\widetilde Q_v)$ which are related by
the field redefinition Eq.~(\ref{eq53})
\begin{eqnarray}
{\cal L}(Q_v) \equiv  \widetilde  {\cal L}\left(Q_v + \frac{d}{m_Q} (i D \cdot v) Q_v\right)\,.
\end{eqnarray}
The current $\widetilde j^0$ for $\widetilde {\cal L}(\widetilde Q_v)$ is obtained by making the symmetry transformation
\begin{eqnarray}
\widetilde Q_v & \to & e^{i \alpha} \widetilde Q_v
\label{eq55}
\end{eqnarray}
with $\alpha$ a function of time, and picking out the $\dot \alpha$ term,
\begin{eqnarray}
\widetilde  {\cal L}(\widetilde Q_v) &\to& \widetilde  {\cal L}(\widetilde Q_v) -\dot \alpha \widetilde j^0\,.
\end{eqnarray}
Similarly the current $j^0$ for ${\cal L}(Q_v)$ is given by the transformation
\begin{eqnarray}
Q_v & \to & e^{i \alpha} Q_v\,,\nn
{\cal L}(Q_v) &\to& {\cal L}( Q_v) -\dot \alpha j^0\,.
\label{eq57}
\end{eqnarray}
The current $j^0$ is \emph{not} given by substituting Eq.~(\ref{eq53}) for $\widetilde Q_v$ in $ \widetilde j^0$.
The two transformations, Eqs.~(\ref{eq55},\ref{eq57}) are not the same; Eqs.~(\ref{eq57}) gives the
transformation
\begin{eqnarray}
\widetilde Q_v &\to & e^{i \alpha } Q_v + \frac{d}{m_Q} (i D \cdot v)e^{i \alpha} Q_v\nn
&=& e^{i \alpha }\left[\widetilde Q_v - \frac{d}{m_Q} \dot \alpha\, Q_v\right]
\label{eq15}
\end{eqnarray}
which is Eq.~(\ref{eq55}) plus an additional shift in the field. Thus
\begin{eqnarray}
 j^0 &=& \widetilde j^0 + \frac{d}{m_Q} Q_v \frac{\delta \widetilde {\cal L}}{\delta \widetilde Q_v}\,,
\end{eqnarray}
and the two currents differ by a piece proportional to the equations of motion. While the two currents are not the same, their on-shell matrix elements are equal since they differ by equation-of-motion terms.

The full and EFT currents can be computed from the symmetry transformation using the N\"other procedure. In general, the two currents differ, because the fields in the EFT are not directly related to fields in the full theory. Only the symmetry charges, i.e.\ the zero moment-transfer matrix element of $j^0$ agree. One way to see this is to gauge the symmetry current using a background gauge field $\mathcal{A}$. Then the current can be computed  from $\delta S/\delta \mathcal{A}_\mu$. This definition of the current agrees in the full and effective theories, since the two theories are constructed to have the same on-shell scattering amplitude off an external $\mathcal{A}_\mu$ source. 

The current computed from $\delta S/\delta \mathcal{A}_\mu$ can differ from the N\"other current obtained from an infinitesimal symmetry transformation (in the theory with $\mathcal{A}=0$).
In the full theory, the background field enters only in the quark kinetic term $\overline b (i \slashed{D}+ \mathcal{A}) b$. The currents computed using the N\"other method by the field redefinition $b \to e^{i \alpha} b$ in the Lagrangian or from $\delta S/\delta \mathcal{A}_\mu$ are both equal to $j^\mu= \overline b \gamma^\mu b$. However, in the EFT, the two methods give different results. The reason is that the full theory Lagrangian with an external background field can match onto gauge-invariant EFT operators involving the field-strength tensor $\mathcal{F}_{\mu \nu}=\partial_{\mu}\mathcal{A}_\nu-\partial_\nu \mathcal{A}_\mu$
such as $\mathcal{F}_{\mu \nu} O^{\mu \nu}$. The N\"other current is computed from the theory with $\mathcal{A} \to 0$, and has no contribution from this term. However
\begin{eqnarray}
\delta j^\mu = \frac{\delta}{\delta \mathcal{A}_\mu} \int \mathcal{F}_{\mu \nu} O^{\mu \nu} &=&  \partial_\nu O^{\mu \nu}- \partial_\nu O^{\nu \mu},
\end{eqnarray}
so the operator does contribute to the current defined by $\delta S/\delta \mathcal{A}_\mu$ obtained by matching from the full theory. The extra contribution $\delta j^\mu$ is automatically conserved,
\begin{eqnarray}
\partial_\mu \left( \delta j^\mu \right) =0\,,
\end{eqnarray}
\emph{without} use of the equations of motion. Since $\delta j^\mu$ is a gradient, it has zero forward matrix element, and hence does not contribute to the charges.

We will assume that the HQET Lagrangian has the form Eq.~(\ref{eq1}). It is worth noting that even if one starts with an EFT with no higher-order time derivatives (such as Eq.~(\ref{eq1})), such time-derivatives can be introduced by the renormalization counterterms, so that a further field redefinition is necessary to put the Lagrangian back into canonical form. One way to avoid this is to only renormalize $S$-matrix elements, which are independent of the choice of field. 

The EFT N\"other current is $\overline Q_v Q_v$. For inclusive decays, we need only the forward matrix element of this operator, so we can use the result that it is unity to all orders in $1/m_Q$.

\section{Reparametrization Invariance}\label{sec:rpi}

In this section, we review some results on RPI derived in Refs.~~\cite{Luke:1992cs,Manohar:1997qy} which are required for our analysis. The RPI transformation $k \to k - l$, $v \to w=v + l/m_Q$  on heavy quark fields was given in Ref.~\cite{Luke:1992cs}. One defines the RPI covariant field $\Psi_v$ in terms of the field $\psi_v$ satisfying $\slashed{v}\psi_v=\psi_v$,
\begin{eqnarray}
\Psi_v(x) &=& \Lambda(u,v) \psi_v(x)\,,\nn
\Lambda(u,v) &=& \frac{1 + \slashed{u} \slashed{v}}{\sqrt{2 \left(1 +u \cdot v \right)}} \,,\nn
u^\mu &=& \frac{v^\mu + i D^\mu/m_Q}{\abs{v^\mu + i D^\mu/m_Q}}\,,
\label{eq38}
\end{eqnarray}
which transforms covariantly under RPI as
\begin{eqnarray}
\Psi_w(x) &=& e^{i l \cdot x} \Psi_v(x)\,.
\label{eq39}
\end{eqnarray}
$\psi_v$ is related to the field $Q_v$ used earlier by a field redefinition, Eq.~(\ref{hv}), as discussed below.
One picks a standard operator ordering for $\Lambda$. Different choices of operator ordering correspond to a field redefinition of the field $\psi_v$. RPI invariant operators can then be constructed from $\Psi_v$, e.g.\
\begin{eqnarray}
\overline \Psi_v \Psi_v,\qquad \overline \Psi_v \left(v+\frac{iD}{m_Q}\right)^2 \Psi_v, \qquad \text{etc.}
\end{eqnarray}

The RPI invariant Lagrangian is constructed by writing down all possible RPI invariant Lorentz invariant operators. The RPI invariant operators written in terms $\psi_v$ contain higher order time-derivatives, and the RPI invariant Lagrangian constructed from Eq.~(\ref{eq38}) is not in the canonical form Eq.~(\ref{eq1}), when written in terms of $\psi_v$. As noted earlier, it is convenient to make a field redefinition so that higher order time-derivatives are removed from the Lagrangian. The required field redefinition is
\begin{eqnarray}
\psi_v &=& \Biggl[1-
\frac{g }{16 m_Q^2} \sigma_{\alpha \beta}G^{\alpha \beta}-\frac{1}{4 m_Q^2}D_\perp^2
-\frac{g}{8m_Q^3}\sigma_{\alpha \beta}G^{\alpha \beta}(i v \cdot D)\nn
&&
+ \frac{1}{4m_Q^3} \slashed{D}_\perp (i v \cdot D)  \slashed{D}_\perp \Biggr]Q_v\,.
\label{hv}
\end{eqnarray}
The basic RPI invariant operators after the field redefinition are
\begin{eqnarray}
&& \overline \Psi_v  \Psi_v = \overline Q_v Q_v +\frac{O_2}{m_Q}+\frac{O_F}{m_Q}+\frac{2O_D}{m_Q} +\frac{2O_S}{m_Q}\nn
&&\qquad +  \frac{1}{4m_Q^3} 
\overline Q_v \left\{i v \cdot D,D_\perp^2\right\}Q_v\nn
&&\qquad +\frac{g }{8 m_Q^3}\overline Q_v \left\{iv \cdot D, \sigma_{\alpha \beta}G^{\alpha \beta}\right\}Q_v
+\mathcal{O}\left( \frac{1}{m_Q^4}\right)\,,\nonumber
\end{eqnarray}
\begin{eqnarray}
&& \overline \Psi_v \left( \slashed{p}-m_Q \right) \Psi_v = \overline Q_v (i v \cdot D) Q_v +O_2+O_F\nn
&&\qquad+O_D+O_S+\mathcal{O}\left( \frac{1}{m_Q^3}\right)\,,\nonumber
\end{eqnarray}
\begin{eqnarray}
&& -\frac{g}{4m_Q} \overline \Psi_v \sigma_{\alpha \beta} G^{\alpha \beta} \Psi_v 
= O_F + 2 O_S +2 O_D+\mathcal{O}\left( \frac{1}{m_Q^3}\right)\,,\nonumber
\end{eqnarray}
\begin{eqnarray}
&& \frac{i}{16m_Q^3}\overline \Psi_v \left\{p_\alpha, \left[p_\beta,g G^{\alpha \beta}\right] \right\}\Psi_v 
= O_D+\mathcal{O}\left( \frac{1}{m_Q^3}\right)\,,\nn
\label{eq101}
\end{eqnarray}
where $p=m_Q v + i D$. Other RPI invariant operators are linear combinations of these operators.

The most general Lagrangian is a linear combination of the RPI invariant combinations in Eqs.~(\ref{eq101}), with a kinetic term of unit coefficient, and no residual mass term. This gives (for the heavy quark fermion bilinear terms)
\begin{eqnarray}
L &=& \left[\overline Q_v (i v \cdot D) Q_v +O_2+O_F+O_D+O_S\right]\nn
&&+c_1\left[O_F + 2 O_S +2 O_D\right]+c_2\left[O_D\right]+\mathcal{O}\left( \frac{1}{m_Q^3}\right)\nn
\label{eq46}
\end{eqnarray}
with arbitrary coefficients $c_{1,2}$, and leads to the RPI relations~\cite{Luke:1992cs,Manohar:1997qy}
\begin{eqnarray}
c_2 &=& 1\,,\nn
c_S &=& 2 c_F-1\,,
\label{cscf}
\end{eqnarray}
for the coefficients of the HQET Lagrangian Eq.~(\ref{eq1}). The relations hold to all orders in $\alpha_s$, and for arbitrary renormalization scale $\mu$. This implies that $c_2$ and $O_2$ have no anomalous dimension, so that $\lambda_1$ is $\mu$-independent~\cite{Luke:1992cs}. The renormalization group evolution equations for $c_S$ and $c_F$ maintain the relation $c_S=2c_F-1$ at all scales~\cite{Bauer:1997gs} (see the appendix).

The matrix element of $\overline Q_v Q_v$ is unity, from the analysis in Sec.~\ref{sec:currents}. It is incorrect to assume that the full theory current is the RPI invariant operator $\overline \Psi_v \gamma^\mu \Psi_v$. $\Psi_v$ is not the full theory field. The  full theory current matches on to the most general RPI invariant operator that transforms as a four-vector, and so has additional terms. The matrix element of $\overline \Psi_v \gamma^\mu \Psi_v$ is not unity.

\section{Total Decay Rate}\label{sec:decayrate}

The total inclusive decay rate can be written as the forward matrix element of a local decay operator $\widehat \Gamma$,
\begin{eqnarray}
\Gamma &=& \braket{\bar B^{(*)}\!\!,p|\, \widehat \Gamma\, | \bar B^{(*)}\!\!,p}
\end{eqnarray}
between full theory states~\cite{Bigi:1992su}. This follows by using the optical theorem to write the total decay rate as the imaginary part of the forward scattering amplitude, and then performing an operator product expansion (OPE). More details are given in Sec.~\ref{sec:spectra}. 

The decay operator $\widehat \Gamma$ is a local scalar operator, and has no knowledge of any kinematic variables, since we have integrated over all final states. It can therefore be written as a linear combination of the RPI invariant operators in Eq.~(\ref{eq101}),
\begin{eqnarray}
\frac{\widehat \Gamma}{\Gamma_0} &=& d_1 \biggl[ \overline Q_v Q_v +\frac{O_2}{m_Q}+\frac{O_F}{m_Q}+\frac{2O_D}{m_Q} +\frac{2O_S}{m_Q}\nn
&&+  \frac{1}{4m_Q^3} 
\overline Q_v \left\{i v \cdot D,D_\perp^2\right\}Q_v\nn
&& +\frac{g }{8 m_Q^3}\overline Q_v \left\{iv \cdot D, \sigma_{\alpha \beta}G^{\alpha \beta}\right\}Q_v\biggr]\nn
&&+\frac{d_2}{m_Q} \left[\overline Q_v (i v \cdot D) Q_v +O_2+O_F+O_D+O_S\right]\nn
&&+\frac{d_3}{m_Q} \left[O_F + 2 O_S +2 O_D\right]+\frac{d_4}{m_Q}\left[O_D\right]\nn
&&+\sum_i \frac{d_i^{(2)}}{m_Q^3} O_i^{(2)}+\mathcal{O}\left( \frac{1}{m_Q^4}\right)
\label{eq29}
\end{eqnarray}
where  $\Gamma_0= \abs{V_{qQ}}^2 G_F^2 m_Q^5/(192 \pi^3)$ for inclusive $Q \to X_q \ell \nu$ decay is the conventional normalization which has been factored out, so that $d_i$ are dimensionless coefficients
which can depend on $\alpha_s$, $r=m_q^2/m_Q^2$, and $\log m_Q/\mu$. It is important that the Lagrangian Eq.~(\ref{eq46}) and the decay width Eq.~(\ref{eq29}) are written in terms of the \emph{same} field $Q_v$.

We can now take the on-shell matrix element of $\widehat \Gamma$ between full theory states. The $iv \cdot D$ terms can be eliminated using the equation of motion of the Lagrangian Eq.~(\ref{eq46}), so that 
$\widehat \Gamma$ can be replaced by
\begin{eqnarray}
\frac{\widehat \Gamma}{\Gamma_0}&\to& d_1 \Bigl[ \overline Q_v Q_v +\frac{O_2}{m_Q}+\frac{O_F}{m_Q}+\frac{2O_D}{m_Q} +\frac{2O_S}{m_Q}\Bigr]\nn
&&+\frac{d_3}{m_Q} \left[O_F + 2 O_S +2 O_D\right]+\frac{d_4}{m_Q}\left[O_D\right]\nn
&&+\sum_i \frac{d_i^{(2)}}{m_Q^3}O_i^{(2)}+\mathcal{O}\left( \frac{1}{m_Q^4}\right)
\label{eq30}
\end{eqnarray}
between on-shell states. The $d_2$ contribution has been absorbed into the remaining $d_i$ by
$d_3 + d_2(1-c_F) \to d_3$, $d_4 +d_2(c_F-c_D) \to d_4$, $d_i^{(2)}-d_2 c_i^{(2)} \to d_i^{(2)}$ which is possible because of Eq.~(\ref{cscf}). The analysis of Sec.~\ref{sec:currents} shows that the matrix element of $\overline Q_v Q_v$ is unity to all orders in $1/m_Q$ and $\alpha_s$. This result is crucial, because otherwise there would be unknown terms in Eq.~(\ref{eq31}) beyond tree-level.
Using Eqs.~(\ref{eq5}), and expanding out the full theory states in terms of HQET states using time-ordered products with $1/m_Q$ suppressed terms in the Lagrangian (see Ref.~\cite{Gremm:1996df}) gives
\begin{eqnarray}
\frac{\widehat \Gamma}{\Gamma_0}&=& d_1 \left[ 1 +\frac{1}{2m_Q^2}\left(
\lambda_1+\frac{\tau_1+c_F d_H \tau_2}{m_Q}\right) \right]\nn
&&\hspace{-1cm} +d_2^\prime  \left[ \frac{1}{2m_Q^2}\left(d_H \lambda_2 + \frac{c_F \tau_3 +d_H\left( \tau_2+c_F \tau_4\right)}{ m_Q}\right)-\frac{d_H \rho_2}{2m_Q^3}\right]\nn
&& -d_3^\prime \frac{\rho_1}{2m_Q^3}+\sum_i \frac{d_i^{(2)}}{m_Q^3} \vev{O_i^{(2)}}+\mathcal{O}\left( \frac{1}{m_Q^4}\right)
\label{eq31}
\end{eqnarray}
where $d_2^\prime=d_1+d_3$ and $d_3^\prime=d_1+d_3-d_4/2$, and the time-ordered products are defined by
\begin{eqnarray}
&&i\braket{\bar B^{(*)}\!\!,v|T\left\{ O_2 (0) \int  \rd^4 x\, O_2(x) \right\}|\bar B^{(*)}\!\!,v} = \frac{\tau_1}{2m_Q^2}\,,\nonumber
\end{eqnarray}
\begin{eqnarray}
&& i\braket{\bar B^{(*)}\!\!,v|T\left\{ O_2 (0) \int  \rd^4 x\, O_F(x) \right\}|\bar B^{(*)}\!\!,v} = \frac{d_H \tau_2}{2m_Q^2}\,,\nonumber
\end{eqnarray}
\begin{eqnarray}
&& i\braket{\bar B^{(*)}\!\!,v|T\left\{ O_F (0) \int  \rd^4 x\, O_2(x) \right\}|\bar B^{(*)}\!\!,v} = \frac{d_H \tau_2}{2m_Q^2}\,,\nonumber
\end{eqnarray}
\begin{eqnarray}
&& i\braket{\bar B^{(*)}\!\!,v|T\left\{ O_F (0) \int  \rd^4 x\, O_F(x) \right\}|\bar B^{(*)}\!\!,v} = \frac{\tau_3}{2m_Q^2}+\frac{d_H \widetilde \tau_4}{2m_Q^2}\,.\nn
\label{tprod}
\end{eqnarray}
The time-ordered products need renormalization beyond that in the operators $O_{2,F}$, and $\tau_i$ are the finite matrix element of the renormalized time-ordered products.

The time-ordered products $\mathcal{T}_{1-4}$ in Ref.~\cite{Gremm:1996df} are defined using the tree-level Lagrangian with $c_F=1$, and are related to $\tau_{1-4}$ by
\begin{eqnarray}
\mathcal{T}_1 &=& \tau_1\,,\nn
\mathcal{T}_2 &=& c_F \tau_2\,,\nn
\mathcal{T}_3 &=& c_F \tau_3\,,\nn
\mathcal{T}_4 &=& \tau_2 + c_F \tau_4\,,
\label{eq33}
\end{eqnarray}
with $c_F=1$. 
The matrix elements of the operators $O_{2,F}$ including $1/m_Q$ corrections to the states are given by
\begin{eqnarray}
\lambda_1 &\to& \lambda_1 + \frac{\tau_1+d_H c_F \tau_2}{2m_Q}\,,\nn
d_H \lambda_2 &\to& d_H \lambda_2 + \frac{d_H \tau_2+c_F \tau_3+d_H  c_F\widetilde \tau_4}{2m_Q}\,,
\label{a4}
\end{eqnarray}
in Eq.~(\ref{eq5}).

Writing the decay rate to order $1/m_Q^3$ as
\begin{eqnarray}
\Gamma &=& \Gamma_0 \Biggl[ f_0 + f_{\lambda_1} \frac{\lambda_1}{m_Q^2} + f_{\lambda_2}   \frac{d_H \lambda_2}{m_Q^2}\nn
&&+ f_{\tau_1}   \frac{\tau_1}{m_Q^3}+ f_{\tau_2}   \frac{d_H \tau_2}{m_Q^3}+ f_{\tau_3}   \frac{\tau_3}{m_Q^3}+ f_{\tau_4}   \frac{d_H \tau_4}{m_Q^3}\nn
&&+ f_{\rho_1} \frac{\rho_1 }{m_Q^3}+ f_{\rho_2} \frac{d_H\rho_2}{m_Q^3} +f_i^{(2)} \frac{\vev{O_i^{(2)}}}{m_Q^3}\Biggr]
\label{eq34}
\end{eqnarray}
where $f_i$ depend on $\alpha_s$, $r$, and $\log m_Q/\mu$, we see that Eq.~(\ref{eq31}) implies the relations
\begin{eqnarray}
f_{\tau_1} &=& f_{\lambda_1} = \frac12 f_0\,,\nn
f_{\tau_2} &=&  c_F f_{\lambda_1}+f_{\lambda_2}\,,\nn
f_{\tau_3} &=& f_{\tau_4}= c_F f_{\lambda_2}\,,\nn
f_{\rho_2} &=& - f_{\lambda_2}\,.
\label{relns}
\end{eqnarray}
The relation  $f_{\lambda_1} = f_0/2$ between the $\lambda_1$ term and the leading order term has been known for a long time~\cite{Bigi:1992su,Manohar:1993qn}. This relation is valid at any $\mu$ since $\lambda_1$ is $\mu$ independent. The four-quark terms $f_i^{(2)}$ are present, and are needed for the decay rate to be $\mu$-independent. Under renormalization group evolution $O_D$, $O_S$, the time-ordered products and the four-quark operators mix~\cite{Bauer:1997gs}. 

The expression for the hadron masses to order $\lqcd^3/m_Q^2$ is given in the appendix. The appendix also discusses the $\mu$-independence of the mass, and the connection between $O_D$ and the four-quark operators $O^{(2)}_i$. This is a simpler analysis than that for the hadronic tensor, and illustrates some of the features due to operator mixing and time-ordered products, as well as the interplay between RPI and renormalization.

The tree-level decay rate to $1/m_Q^3$ of Ref.~\cite{Gremm:1996df} is
\begin{eqnarray}
\Gamma &=& \Gamma_0 \biggl[ f_0(r) + f_{\lambda_1} (r) \frac{1}{m_Q^2} \left(\lambda_1 
+ \frac{\mathcal{T}_1+3\mathcal{T}_2}{m_Q}\right)\nn
&& + f_{\lambda_2} (r)   \frac{1}{m_Q^2}\left( \lambda_2 + \frac{\mathcal{T}_3+3\mathcal{T}_4}{3m_Q} \right)\nn
&&+ \frac{1}{m_Q^3}f_{\rho_1}(r) \rho_1 +\frac{1}{m_Q^3}f_{\rho_2}(r)\rho_2 \biggr]\,,
\end{eqnarray}
where
\begin{eqnarray}
f_0 &=&  1 - 8 r +8 r^3-r^4-12 r^2 \log r \,,\nn
f_{\lambda_1} &=&  \frac12 - 4 r +4 r^3-\frac12 r^4-6 r^2 \log r \,,\nn
f_{\lambda_2} &=&  -\frac92+12 r -18 r^2 \log r -36 r^2 +36 r^3 -\frac{15}{2}  r^4\,,\nn
f_{\rho_1} &=&  \frac{77}{6}+ 8 \log r -\frac{44}{3}r+10r^2-\frac43 r^3-\frac56 r^4 \,,\nn
f_{\rho_2} &=& \frac92-12 r +18 r^2 \log r + 36 r^2 -36 r^3 +\frac{15}{2}  r^4\,,
\end{eqnarray}
and satisfies Eq.~(\ref{relns}) using Eq.~(\ref{eq33}) and $c_F=1$. The $\lambda_1$  and $\rho_2$ contributions to the total decay rate occur in the combinations $1+\lambda_1/(2m_Q^2)$, and $\lambda_2-\rho_2/m_Q$.

\section{Inclusive Decay Spectra}\label{sec:spectra}

The inclusive decay spectra for semileptonic $B \to X_c e \overline \nu_e$ were computed in Ref.~\cite{Manohar:1993qn} to order $1/m_Q^2$. The $1/m_Q^3$ corrections were given in Ref.~~\cite{Gremm:1996df}. The inclusive decay spectra can be determined from the hadronic tensor $W_{\mu\nu}$, where
\begin{eqnarray}
W_{\mu \nu} &=& -\frac{1}{\pi}\text{Im}\, T_{\mu \nu}
\end{eqnarray}
and
\begin{eqnarray}
T_{\mu \nu} &=& -i \int \rd^4 x e^{-i q \cdot x} \frac{\braket{\bar B| T \left\{ J_{L \mu}^\dagger(x) J_{L\nu}(0)\right\} |\bar B}}{2 m_B}\,.
\end{eqnarray}
$W_{\mu\nu}$ has the tensor decomposition
\begin{eqnarray}
W_{\mu \nu} &=& - g_{\mu \nu}W_1+v_\mu v_\nu W_2 + i W_3 \epsilon_{\mu \nu \alpha \beta}q^\alpha
v^\beta + W_4 q_\mu q_\nu \nn
&&+W_5\left(q_\mu v_\nu+q_\nu v_\mu\right)
\end{eqnarray}
where the invariant tensors $W_i$ are functions of the kinematic variables $q^2$, $q \cdot v$.

RPI constraints on $T^{\mu\nu}$ can be determined by writing $T^{\mu\nu}$ as a linear combination of RPI invariant operators. The algebra is considerably more complicated than for the total decay rate, because the decay tensor has two indices, and can depend on the momentum $q$. Only the final results are given here. The decay tensor is written in an expansion similar to Eq.~(\ref{eq34}),
\begin{eqnarray}
T_{\mu\nu} &=& T_{\mu\nu}^{(0)} + T_{\mu\nu}^{(\lambda_1)} \frac{\lambda_1}{m_Q^2} + T_{\mu\nu}^{(\lambda_2)}  \frac{d_H \lambda_2}{m_Q^2}\nn
&&+ T_{\mu\nu}^{(\tau_1)} \frac{\tau_1}{m_Q^3}+ T_{\mu\nu}^{(\tau_2)}   \frac{d_H \tau_2}{m_Q^3}+ T_{\mu\nu}^{(\tau_3)}   \frac{\tau_3}{m_Q^3}+T_{\mu\nu}^{(\tau_4)}   \frac{d_H \tau_4}{m_Q^3}\nn
&&+ T_{\mu\nu}^{(\rho_1)} \frac{\rho_1 }{m_Q^3}+  T_{\mu\nu}^{(\rho_1)} \frac{d_H\rho_2}{m_Q^3} + T_{\mu\nu}^{(2),i}
\frac{ \vev{O_i^{(2)}} }{m_Q^3}\,.
\label{eq41}
\end{eqnarray}
The time-ordered products are related to the $\lambda_{1,2}$ terms, 
\begin{eqnarray}
T_{\mu\nu}^{(\tau_1)} &=& T_{\mu\nu}^{(\lambda_1)} \,,\nn
T_{\mu\nu}^{(\tau_2)} &=&  c_F T_{\mu\nu}^{(\lambda_1)} + T_{\mu\nu}^{(\lambda_2)} \,,\nn
T_{\mu\nu}^{(\tau_3)}  &=& T_{\mu\nu}^{(\tau_4)} = c_F T_{\mu\nu}^{(\lambda_2)}\,.
\label{eq42}
\end{eqnarray}

The non-trivial relations are for the $\lambda_1$ and $\rho_2$ terms. The RPI constraint on the $\lambda_1$ term is
\begin{eqnarray}
T_{\mu \nu}^{(\lambda_1)} &=& \frac12 T_{\mu \nu}^{(0)} -\frac12  v^\alpha \frac{\partial T_{\mu \nu}^{(0)}}{\partial v^\alpha}+\frac16
g_\perp^{\alpha \beta}  \frac{\partial^2 T_{\mu \nu}^{(0)}}{\partial v^\alpha v^\beta} \,.
\label{eq43}
\end{eqnarray}
The $v$ derivatives act on the explicit factors of $v^\mu$ and $v^\nu$ in the tensor decomposition of $T_{\mu \nu}$, as well as on the $v$ dependence of the kinematic variable $q \cdot v$. The velocity four-vector $v$ satisfies $v \cdot v=1$. This equation is consistent with
Eq.~(\ref{eq43}) because
\begin{eqnarray}
0 &=&  \left(-\frac12  v^\alpha \frac{\partial }{\partial v^\alpha}+\frac16
g_\perp^{\alpha \beta}  \frac{\partial^2 }{\partial v^\alpha v^\beta}\right)\left(v \cdot v\right) \,.
\label{44}
\end{eqnarray}
The two terms are non-zero, but the sum vanishes. In terms of the invariant tensors $T_i$, Eq.~(\ref{eq43}) gives the relations
\begin{eqnarray}
T^{(\lambda_1)}_i &=& \frac12 T^{(0)}_i-\frac12\hat q \cdot v \frac{\partial}{\partial (\hat q \cdot v)}T_i^{(0)}\nn
&&+ \frac 16 \left[ \hat q^2 -  (\hat q \cdot v)^2\right]
\frac{\partial^2}{\partial (\hat q \cdot v)^2}T^{(0)}_i+\delta T_i\,,
\end{eqnarray}
\begin{eqnarray}
\delta T_1 &=& -\frac 13 T_2^{(0)} \,,
\end{eqnarray}
\begin{eqnarray}
\delta T_2 &=& -\frac43 T_2^{(0)} - \frac23 \hat q \cdot v \frac{\partial T_2^{(0)}}{\partial (\hat q \cdot v)}\,,
\end{eqnarray}
\begin{eqnarray}
\delta T_3 &=& -\frac12 T_3^{(0)} - \frac13\hat q \cdot v \frac{\partial T_3^{(0)}}{\partial (\hat q \cdot v)}\,,
\end{eqnarray}
\begin{eqnarray}
\delta T_4 &=&  \frac23  \frac{\partial T_5^{(0)}}{\partial (\hat q \cdot v)}\,,
\end{eqnarray}
\begin{eqnarray}
\delta T_5 &=& -\frac12T_5 +\frac13 \frac{\partial T_2^{(0)}}{\partial (\hat q \cdot v)} -\frac13 \hat q \cdot v \frac{\partial T_5^{(0)}}{\partial (\hat q \cdot v)}\,,
\label{eq82}
\end{eqnarray}
where $\hat q=q/m_Q$. These relations are equivalent to the relations found in Sec.~VI of Ref.~\cite{Manohar:1993qn}.

The differential decay rate is given by multiplying the hadronic tensor $T^{\mu\nu}$ with the leptonic tensor $L_{\mu\nu}$, which is a function of $k_e$ and $k_\nu$, the four-momenta of the electron and neutrino. If we integrate this over phase space with a weight function that is independent of $v$, then the $v$ derivatives in Eq.~(\ref{eq43}) give zero after integrating by parts, and the integral is proportional to $1+\lambda_1/(2m_Q^2)$. This is true for the total decay rate, and also for the $q^2$ spectrum. It is not true for the electron energy spectrum, since the electron energy is $E_e=k_e \cdot v$.
 
The RPI constraint on the $\lambda_2$ and $\rho_2$ terms in $T^{\mu\nu}$ is that they must occur in the linear combination
\begin{eqnarray}
&& \left(\frac{\rho_2}{m_Q} -\lambda _2 \right)\left(g^{\alpha \beta}-2v^\alpha v^\beta\right)Y^{\mu\nu}_{\alpha \beta} - \frac{\rho_2}{m_Q}
 \frac{\partial }{\partial_\perp v^\lambda}\Bigl[ v^\beta g_\perp^{\alpha \lambda} Y^{\mu\nu}_{\alpha \beta} \Bigr]\nn
\label{eq128a}
\end{eqnarray}
where $\partial_\perp$ is defined by
\begin{eqnarray}
\frac{\partial f}{\partial_\perp v^\alpha} &\equiv& g_\perp^{\alpha \beta} \frac{\partial f}{\partial v^\beta} \,,
\end{eqnarray}
The $\perp$ derivative is useful because the chain rule $\partial_\perp(f g) = f (\partial_\perp g)+ (\partial_\perp f)g$ is compatible with the constraint $v \cdot v=1$. $Y^{\mu \nu}_{\alpha \beta}$ is a tensor which is a function of $q$ and $v$.  The explicit form of
$Y^{\mu \nu}_{\alpha \beta}$ is not particularly enlightening.

The last term in Eq.~(\ref{eq128a}) is
\begin{eqnarray}
&&  - \frac{\rho_2}{m_Q}\left(g^{\lambda \sigma} - v^\lambda v^\sigma\right)
 \frac{\partial }{\partial v^\sigma}\Bigl[ v^\beta g_\perp^{\alpha \lambda} Y^{\mu\nu}_{\alpha \beta} \Bigr]\nn
&=&  - \frac{\rho_2}{m_Q}\frac{\partial}{\partial v^\sigma}\left\{ \left(g^{\lambda \sigma} - v^\lambda v^\sigma\right)\Bigl[ v^\beta g_\perp^{\alpha \lambda} Y^{\mu\nu}_{\alpha \beta} \Bigr]
 \right\}\,.
\label{eq129}
\end{eqnarray}
The integral of the last term over any weight function independent of $v$ vanishes. In particular, it vanishes for the total rate and the $q^2$ spectrum, so both must be proportional to $\lambda_2-\rho_2/m_Q$, which agrees with Eq.~(\ref{relns}).

\section{$\bar B \to X_s \gamma$, $\bar B \to X_s l^+ l^-$}\label{sec:bsg}

The RPI constraints have been derived for the $\bar B \to X_q e \overline \nu$ inclusive decay spectra. The same analysis goes through unchanged for inclusive $\bar B \to X_s l^+ l^-$ and $\bar B \to X_s \gamma$ decays. The $\bar B \to X_s l^+ l^-$ inclusive decay rate is described by a hadronic tensor $T^{\mu\nu}$ which satisfies the same RPI relations as $T^{\mu\nu}$ for inclusive semileptonic decay. $\bar B \to X_s \gamma$ decay is described by a hadronic tensor $T(q)$ with no indices. It satisfies the RPI constraints Eq.~(\ref{eq42}), Eq.~(\ref{eq43}) and Eq.~(\ref{eq128a}) with the $\mu$ and $\nu$ indices deleted. Eq.~(\ref{eq82}) reduces to the equation
\begin{eqnarray}
T^{(\lambda_1)} &=& \frac12 T^{(0)}-\frac12\hat q \cdot v \frac{\partial}{\partial (\hat q \cdot v)}T^{(0)}\nn
&&+ \frac 16 \left[ \hat q^2 -  (\hat q \cdot v)^2\right]
\frac{\partial^2}{\partial (\hat q \cdot v)^2}T^{(0)}\,.
\label{eq48}
\end{eqnarray}
These RPI relations for $\bar B \to X_s \gamma$ and $\bar B \to X_s l^+ l^-$ are satisfied by the results in Refs.~\cite{Falk:1993dh,Ali:1996bm,Bauer:1997fe,Bauer:1999kf} for the decay tensors to order $1/m_Q^3$.

\acknowledgments

I would like to thank Z.~Ligeti, M.E.~Luke, and F.~Tackmann for helpful discussions.

\begin{appendix}

\section{Hadron Masses}\label{sec:masses}

Gremm and Kapustin~\cite{Gremm:1996df} computed the hadron masses to order $1/m_Q^3$ at tree-level.
Their result can be easily generalized to include radiative corrections by using the HQET Lagrangian Eq.~(\ref{eq1}), with coefficients $c_{F,S,D}$. The masses are (with $c_2=1$)
\begin{eqnarray}
M_H &=& m_Q +\overline\Lambda  -\frac{\lambda_1 + d_H c_F \lambda_2 }{2m_Q}\nn
&& +\frac{c_D \rho_1-\tau_1-c_F^2 \tau_3 + d_H\left(c_S \rho_2 - 2 c_F \tau_2
-c_F^2 \tau_4\right)}{4m_Q^2}\nn
&&+  \frac{1}{m_Q^2}\sum_i c_i^{(2)}\Delta M^{(2)}_i\,.
\label{a1}
\end{eqnarray}
The order $\lqcd$ term $\overline \Lambda$ is defined by
\begin{eqnarray}
\braket{\bar B^{(*)}\!\!,v|H_0 |\bar B^{(*)}\!\!,v} &=& -\overline \Lambda\,,
\end{eqnarray}
where $H_0$ is the $m \to \infty$ Hamiltonian constructed from the $m \to \infty$ Lagrangian ${\cal L}_l
+ \overline Q_v (iD \cdot v)Q_v$. The order $\lqcd^3$ terms $\Delta M^{(2)}_i$ are given by
\begin{eqnarray}
\braket{\bar B^{(*)}\!\!,v|O_i^{(2)}|\bar B^{(*)}\!\!,v} &=& -\Delta M^{(2)}_i\,.
\end{eqnarray}

The renormalization group evolution of the terms in Eq.~(\ref{a1}) is given in Ref.~\cite{Bauer:1997gs}.
In the notation here, $O_2=O_k$, $O_F=-O_m$, and the time-ordered products in Eq.~(\ref{tprod}) are
$2O_{kk}$, $-O_{km}$, $-O_{km}$ and $2O_{mm}$, respectively. One element of $\left\{ O_i^{(2)}/m_Q^2\right\}$ is the four-quark operator
\begin{eqnarray}
  O_1^{hl}&=&\frac{g^2}{8m_Q^2}\sum_i\overline Q_v
  T^A Q_v\ \bar{q}_i\slashed{v} T^A q_i,
\end{eqnarray}
where the sum is over light quark flavors. The gluon equation of motion reduces to
\begin{eqnarray}
O_D &=& O_1^{hl}
\label{eom}
\end{eqnarray}
in the single heavy quark sector. Thus $\rho_1$ and $\Delta M^{hl}_1$ are related,
\begin{eqnarray}
-\frac{\rho_1}{4m_Q^2} &=&\vev{O_D} = \vev{O_1^{hl}} = \frac{1}{m_Q^2}\Delta M^{hl}_1\,.
\end{eqnarray}

The split of the Lagrangian into $O_D$ and $O_1^{hl}$ is not unique since the two operators are related by the equations of motion. The renormalization group evolution is also not unique. The $1/m_Q^2$ operator $ O$ can have an RG evolution
\begin{eqnarray}
\mu\frac{\rd}{\rd \mu}O &=& * + c \left(O_D - O_1^{hl}\right)\,,
\end{eqnarray}
i.e.\ the anomalous dimension can have a piece proportional to the equation of motion. $c$ is not unique, and depends, for example, on the choice of gauge. The equation of motion Eq.~(\ref{eom}) is preserved under renormalization. This point is discussed in more detail in Ref.~\cite{Bauer:1997gs}. 

One can eliminate $O_1^{hl}$ from the operator basis using Eq.~(\ref{eq5}). The RG equations in this reduced basis are~\cite{Bauer:1997gs}:\footnote{$c_F$ is the coefficient of the Fermi term, and $C_F=4/3$ is the Casimir of the fundamental representation.}
\begin{eqnarray}
\mu \frac{\rm d}{\rm d\mu} c_2 &=& 0 \,,\nn
\mu \frac{\rm d}{\rm d\mu} c_F &=& \frac{\alpha_s}{4\pi} \left(2 C_A c_F\right) \,,\nn
\mu \frac{\rm d}{\rm d\mu} c_S &=&  \frac{\alpha_s}{4\pi} \left(4 C_A c_2 c_F\right)\,,\nn
\mu \frac{\rm d}{\rm d\mu} c_D &=& \frac{\alpha_s}{4\pi} \left[ \frac{13}{3}C_Ac_D-
\left(\frac{20}{3}C_A+\frac{32}{3}C_F\right)c_2^2-\frac13c_A c_F^2\right] \,,\nn
\label{a8}
\end{eqnarray}
which preserve the RPI relations Eq.~(\ref{cscf}). The mass Eq.~(\ref{a1}) can be seen to be $\mu$-independent using the running of the coefficients and operators given in Ref~\cite{Bauer:1997gs}.
This is (almost) obvious. In most cases (e.g.\ the running of the weak Hamiltonian), the anomalous dimensions of the coefficients and operators are related by
\begin{eqnarray}
\mu \frac{\rm d}{\rm d\mu}  c_i &=& \gamma_{ij} c_j\,,\nn
\mu \frac{\rm d}{\rm d\mu} O_i &=& -\gamma_{ji} O_j\,,
\label{a9}
\end{eqnarray}
so that the effective Lagrangian is $\mu$-independent:
\begin{eqnarray}
\mu \frac{\rm d}{\rm d\mu}  \sum_i c_i O_i &=& 0\,.
\end{eqnarray}
The renormalization group equations for an EFT are non-linear, the running of $1/m_Q^2$ coefficients can depend on the product of two $1/m_Q$ coefficients through time-ordered products, e.g.\ the $c_D$ running in Eq.~(\ref{a8}) has a $c_F^2$ contribution. Thus the simple relation Eq.~(\ref{a9}) no longer holds, and there is mixing between local operators and time-ordered products.
The renormalization of the Lagrangian coefficients and operators is determined by making sure that the $S$-matrix has no ultraviolet divergences. The combinatorics between local operator insertions and time-ordered products for the $S$-matrix is the same as for the mass, so that $\mu$-independence of the $S$-matrix also ensures that the mass is $\mu$-independent. It is necessary to include four-quark operators in Eq.~(\ref{a1}) for the mass to be $\mu$-independent. 

The RG analysis of inclusive decay spectra is more involved, since it involves operators other than those in the Lagrangian, and will be discussed elsewhere~\cite{zoltan}. It is essential to include four-quark operator contributions to the hadronic decay tensor for the answer to be $\mu$-independent.

\end{appendix}


\bibliography{rpi}

\end{document}